\documentclass[usenatbib,a4paper,fleqn]{mnras}
\usepackage{newtxtext,newtxmath}
\usepackage[T1]{fontenc}
\usepackage{graphicx}
\usepackage{pdflscape} 
\usepackage{afterpage} 
\usepackage{pdfendlink}

\hypersetup{pdfauthor={Y. C. Perrott},
               pdftitle={AMI-CL~J0300+2613: a Galactic anomalous-microwave-emission ring masquerading as a galaxy cluster},
               pdfkeywords={dust, extinction, galaxies: clusters: individual: AMI-CL J0300+2613, infrared: ISM, radio continuum: ISM},
               bookmarksnumbered=true}

\graphicspath{{./figs/}}

\newcommand{\AMIAC}{AMI\textsubscript{AC}}
\newcommand{\AMIDC}{AMI\textsubscript{DC}}

\title[An AME ring masquerading as a galaxy cluster]{AMI-CL~J0300+2613: a Galactic anomalous-microwave-emission ring masquerading as a galaxy cluster}
   
    \author[Perrott et~al.]{Yvette~C.~Perrott$^{1}\thanks{Corresponding author: email -- ycp21@mrao.cam.ac.uk}$, Therese~M.~Cantwell$^{2}$, Steve~H.~Carey$^{1}$, Patrick~J.~Elwood$^{1}$, \newauthor
Farhan Feroz$^{1}$, Keith~J.~B.~Grainge$^{2}$, David~A.~Green$^{1}$, Michael~P.~Hobson$^{1}$, \newauthor
Kamran Javid$^{1}$, Terry~Z.~Jin$^{1}$, Guy~G.~Pooley$^{1}$, Nima Razavi-Ghods$^{1}$, \newauthor
Clare Rumsey$^{1}$, Richard~D.~E.~Saunders$^{1}$, Anna~M.~M.~Scaife$^{2}$, Michel~P.~Schammel$^{1}$, \newauthor
Paul~F.~Scott$^{1}$, Timothy~W.~Shimwell$^{3}$, David~J.~Titterington$^{1}$, Elizabeth~M.~Waldram$^{1}$ \\
  $^1$ Astrophysics Group, Cavendish Laboratory, 19 J.~J.~Thomson Avenue, Cambridge CB3 0HE\\
  $^2$ Jodrell Bank Centre for Astrophysics, Alan Turing Building, School of Physics and Astronomy, University of Manchester, \\Oxford Road, Manchester, M13 9PL\\
  $^3$ Leiden University, Rapenburg 70, 2311 EZ Leiden, Netherlands \\
}

\date{Accepted ---; received ---; in original form \today}

\pagerange{\pageref{firstpage}--\pageref{lastpage}}

\pubyear{2017}

\begin{document}
\label{firstpage}

\maketitle

\begin{abstract}
The Arcminute Microkelvin Imager (AMI) carried out a blind survey for galaxy clusters via their Sunyaev--Zel'dovich effect decrements between 2008 and 2011.  The first detection, known as AMI-CL~J0300+2613, has been reobserved with AMI equipped with a new digital correlator with high dynamic range.  The combination of the new AMI data and more recent high-resolution sub-mm and infra-red maps now shows the feature in fact to be a ring of positive dust-correlated Galactic emission, which is likely to be anomalous microwave emission (AME).  If so, this is the first completely blind detection of AME at arcminute scales.
\end{abstract}


\section{Introduction}

The Arcminute Microkelvin Imager (AMI; \citealt{zwart2008}) blind galaxy cluster survey covered $\approx$\,10\,deg$^2$ of the Northern sky, aiming to detect galaxy clusters via the Sunyaev---Zel'dovich (SZ, \citealt{1972CoASP...4..173S}) effect.  Data were taken between 2008 and 2010 on both AMI arrays, the Small Array (SA) to observe the extended cluster emission, and the Large Array (LA) to detect, characterise and subtract the confusing radio point sources to high positional accuracy and sensitivity.  Only the first detection from the survey, known as AMI-CL~J0300+2613, has been published to date in \citet{2012MNRAS.423.1463A}.  This galaxy cluster candidate appeared to be a high-significance, extended, double-peaked SZ source and was also followed up with the Combined Array for Research in Millimeter-wave Astronomy (CARMA; see \citealt{2007ApJ...663..708M} for more details), with which it was detected with lower significance \citep{2013MNRAS.433.2036A}.

Recently, a new digital correlator has been installed on AMI \citep{2017arXiv170704237H}.  Amongst other improvements, this has corrected the point source response which had been subject to baseline-dependent errors introduced by unevenly spaced lags in the analogue-correlator circuit boards and errors in measurement of the longer paths of the analogue path compensation system.  This meant that artefacts were produced near bright sources in \AMIAC\ (we refer to AMI equipped with the analogue correlator as \AMIAC\ and with the new correlator as \AMIDC\ from here on for clarity) maps, which had a complex hour-angle, declination and baseline dependence and were difficult to account for in the data reduction process.  To check whether these artefacts could have induced the discrepancy between the \AMIAC\ and CARMA results, AMI-CL~J0300+2613 was reobserved with \AMIDC.  Here we describe the results of those reobservations.  The paper is organised as follows.  In Sections~\ref{S:obs} to \ref{S:reduction} we describe the observations and data reduction of AMI-CL~J0300+2613 made with \AMIAC\ and \AMIDC.  In Section~\ref{S:analysis} we make qualitative and quantitative comparisons between the \AMIAC\ and \AMIDC\ data and search for alternative explanations for the apparent cluster emission.  In Sections~\ref{S:discussion} and \ref{S:conclusions} we discuss and conclude.

Throughout, we use the colour scale defined in \citet{2011BASI...39..289G}.  Coordinates are in J2000 and we follow the convention $S \propto \nu^{-\alpha}$ for spectral indices.  We assume $H_{0}$ = 70~km~$\rm s^{-1} Mpc^{-1}$ and a concordance $\Lambda$CDM cosmology with $\Omega_{m}$ = 0.3, $\Omega_{\Lambda}$ = 0.7, $\Omega_{k}$ = 0, $\Omega_{b}$ = 0.041, $\omega_{0}$ =$-$1, $\omega_{a}$ = 0 and $\sigma_{8}$ = 0.8. All cluster parameter values are expressed at the redshift of the cluster.

\section{Observations}\label{S:obs}

Characteristics of the two AMI arrays, the SA and LA, are summarised in Table 4 in \citet{2017arXiv170704237H}.  The two arrays are designed to operate in conjunction.  The SA is sensitive to flux on the approximately arcminute scales of the intracluster gas in galaxy clusters at intermediate redshift, but by itself could not separate this extended emission from the confusing emission from compact radio sources in the same line of sight.  In contrast, the LA has longer baselines and is therefore insensitive to the extended galaxy cluster SZ emission (it is `resolved out'), but has higher angular resolution and sensitivity and can therefore determine the positions and fluxes of these compact sources with a high degree of accuracy.  The point source information from the LA is used to subtract these sources from the SA data, leaving only any extended emission that is not visible to the LA.

The analogue correlator operated between $\approx$\,12 -- 18\,GHz, with the passband divided into eight channels of 0.75-GHz bandwidth; the two lowest-frequency channels were discarded due to a combination of low response and the presence of geostationary satellites.  For further details, see \citet{zwart2008}.  The new correlator operates between 13 -- 18\,GHz divided into 4096 channels; this allows the rejection of narrow-band radio-frequency-interference (RFI), making the telescope much more efficient at observing in the presence of RFI.  In addition, the point source response problems have been removed giving a dynamic range of $\sim$\,1000 rather than $\sim$\,100, as well as a slightly improved sensitivity due to the increase in usable bandwidth.  Table~\ref{tab:AMI_obs} summarises the observations carried out to observe AMI-CL~J0300+2613 with \AMIAC\ and \AMIDC; for more information on the survey observations and detection methods see \citet{2012MNRAS.423.1463A}.

\begin{table*}
\centering
\caption{Summary of observations made of AMI-CL~J0300+2613.  The \AMIDC-LA mode is a 61-point hexagonal raster with a greater amount of time spent on the central 19 pointings than the outer ones; both noise levels are indicated in the table, with the lower noise level belonging to the central 19 pointings.}\label{tab:AMI_obs}
\begin{tabular}{lcccccc}\hline
AMI array & Observation mode & Observation dates & Noise level / $\upmu$Jy\,beam$^{-1}$ \\ \hline
\AMIAC-LA & Survey & 2008 Aug -- 2011 May & 44 \\
\AMIDC-LA & 61+19 point raster & 2016 Aug -- 2016 Sep & 32 (110) \\
\AMIAC-SA & Single pointing & 2010 Mar & 67 \\
\AMIDC-SA & Single pointing & 2016 Aug -- 2016 Dec & 57 \\ \hline
\end{tabular}
\end{table*}

\section{Data reduction and mapping}\label{S:reduction}

The \AMIDC\ data were calibrated and imaged in \textsc{casa}\footnote{\url{https://casa.nrao.edu/}}, except for the mapping of mosaics which was performed in \textsc{aips}\footnote{\url{http://aips.nrao.edu/}} due to the current difficulties with defining new primary beam functions in \textsc{casa}.  Primary calibration was performed using a nearby observation of 3C\,286 or 3C\,48, using the \citet{2013ApJS..204...19P} flux density scale along with a correction for the fact that AMI measures \emph{I+Q}, using the polarisation fraction and angle fits from \citet{2013ApJS..206...16P}; this is a $\approx$\,4.5\% correction for 3C\,286 and a $\approx$\,3 -- 5\% correction for 3C\,48, over the AMI band.  The primary calibration observation supplied an instrumental bandpass in both phase and amplitude.  This was applied to the target data, as well as a correction for atmospheric amplitude variations produced by the `rain gauge', which is a noise injection system used to measure the atmospheric noise contribution (see \citealt{zwart2008}).  The nearby bright point source 4C\,28.07 was observed throughout each observation in an interleaved manner and was used to correct for atmospheric and/or instrumental phase drift.  

After narrow-band RFI flagging, the data were binned down to 64 channels to reduce processing time.  The single-pointing SA data were imaged using the \textsc{clean} task, using multi-frequency synthesis with \textit{nterms}=2 which allows for a frequency dependence of the sky brightness.  Multi-scale \textsc{clean} was trialled but did not make a significant difference in the maps so was not used.  For cluster analysis the maps are only used for qualitative purposes; quantitative analysis is carried out in the $uv$-plane to allow for the baseline-dependence of signal from resolved sources.

A large (61-point) raster is necessary to cover the SA field of view with the LA; since the AMI primary beam is not currently modelled within \textsc{casa} we exported the data into $uv$-\textsc{fits} format and imaged using \textsc{imagr} in \textsc{aips}, using the \textsc{flatn} task to combine the raster pointings taking into account the primary beam.

The calibrated \AMIAC\ $uv$-data from \citet{2012MNRAS.423.1463A} were used and re-imaged in an equivalent manner to the \AMIDC\ data.  In the case of the \AMIAC-SA data, two iterations of `flagdata' in `rflag' mode in \textsc{casa} were performed to remove some residual interference striping before re-imaging.

\section{Analysis}\label{S:analysis}

\subsection{Compact radio-source environment}\label{S:point_sources}

\begin{figure}
  \centerline{AMI-LA\hspace{1cm}}
  \begin{center}
    \includegraphics[trim={0cm 1.8cm 0cm 2.2cm},clip=,width=\linewidth]{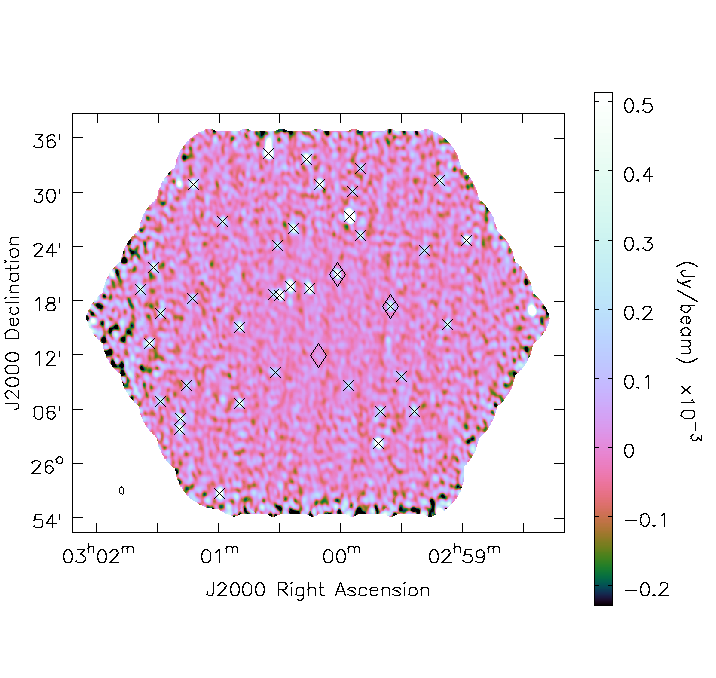}
    \caption{LA map of the compact source environment, made from the combined \AMIAC-LA and \AMIDC-LA datasets.  Crosses mark the positions of the detected sources and diamonds mark the positions of the variable sources discussed in Section~\ref{S:point_sources}. The sources visible at the edges of the map but not marked with crosses are not detected due to failure of the local noise estimation so close to the map edge; they are far enough away from the map centre that failure to subtract them will not affect the SA observations of the target.}
    \label{Fi:comb_map}
  \end{center}
\end{figure}

The \AMIAC-LA and \AMIDC-LA maps were first compared to check for any significant variability or inconsistencies.  Source-finding was carried out down to $4\sigma$ on both maps, using the \textsc{source\_find} software which estimates a local noise level from the map and searches for peaks at a given level of flux density above the noise.  The \textsc{aips} task \textsc{jmfit} was then used to fit a Gaussian model to each source and the deconvolved source size was used to classify each source as point-like or extended, taking into account the signal-to-noise ratio (SNR) of the source; see \citet{2011MNRAS.415.2699A} for more details on the source-finding algorithm and classification scheme.  All sources were found to be point-like, with the exception of AMILA J030035+263425.  On inspection of the maps however, this is clearly two sources quite close together (see e.g.\ the combined map shown in Fig.~\ref{Fi:comb_map}; due to the colour scale the fainter source appears as an extended `tail' to the north of AMILA J030035+263425, which is marked with a cross); the fainter source is very close to the edge of the map and is therefore not detected by the source-finding algorithm, and the Gaussian fit to the brighter source has expanded to include both.  We therefore ignore the extension flag for this source and treat it as point-like, i.e.\ take the peak flux density as the flux estimate.  The fainter source is excluded from the analysis but it is far enough away from the pointing centre that it does not affect the analysis of the SA data.  Sources detected in the maps are listed in Table~\ref{tab:LA_sources}.

\begin{table*}
\centering
\caption{Sources detected in the \AMIAC-LA, \AMIDC-LA, and combined maps.  All sources are point-like and the flux density estimates are the peak flux densities, in mJy\,beam$^{-1}$.  Noise estimates are thermal noise only; when assessing variability we added a 5\% systematic calibration error.  Positions are taken from the combined map or the individual maps if sources are not detected in the combined map.  `Distance' is measured from the pointing centre of the observations.}\label{tab:LA_sources}
\begin{tabular}{lllllllllll}
\hline
Source ID & RA & Dec & \multicolumn{2}{c}{Combined} & \multicolumn{2}{c}{\AMIAC-LA} & \multicolumn{2}{c}{\AMIDC-LA} & Distance & Variable \\
 & & & $S_{\mathrm{peak}}$ & $\Delta S_{\mathrm{peak}}$ & $S_{\mathrm{peak}}$ & $\Delta S_{\mathrm{peak}}$ & $S_{\mathrm{peak}}$ & $\Delta S_{\mathrm{peak}}$ & arcmin \\ \hline
  AMILA J030010+261202 & 03:00:10.43 & +26:12:02.08 &  &  & 0.182 & 0.042 &  &  & 3.3 & V \\
  AMILA J030015+261925 & 03:00:15.17 & +26:19:25.65 & 1.256 & 0.043 & 1.107 & 0.064 & 1.374 & 0.046 & 4.4\\
  AMILA J030024+261941 & 03:00:24.56 & +26:19:41.62 & 1.443 & 0.056 & 1.360 & 0.095 & 1.527 & 0.063 & 5.7\\
  AMILA J030029+261840 & 03:00:29.55 & +26:18:40.31 & 1.526 & 0.100 & 1.508 & 0.091 & 1.602 & 0.128 & 5.8\\
  AMILA J030001+262059 & 03:00:01.34 & +26:20:59.84 & 0.734 & 0.034 & 1.741 & 0.072 & 0.396 & 0.042 & 6.0 & V \\
  AMILA J030032+261849 & 03:00:32.75 & +26:18:49.57 & 0.264 & 0.030 &  &  & 0.323 & 0.050 & 6.5\\
  AMILA J025955+260842 & 02:59:55.81 & +26:08:42.66 & 0.267 & 0.040 &  &  & 0.321 & 0.052 & 7.2\\
  AMILA J030031+261010 & 03:00:31.78 & +26:10:10.22 & 0.217 & 0.037 &  &  &  &  & 7.3\\
  AMILA J025936+261343 & 02:59:36.27 & +26:13:43.51 &  &  &  &  & 0.237 & 0.059 & 7.4\\
  AMILA J025935+261727 & 02:59:35.22 & +26:17:27.00 & 0.461 & 0.034 & 0.689 & 0.057 & 0.340 & 0.045 & 7.8 & V \\
  AMILA J030049+261510 & 03:00:49.44 & +26:15:10.58 & 0.679 & 0.041 & 0.699 & 0.057 & 0.692 & 0.039 & 9.1\\
  AMILA J030031+262411 & 03:00:31.04 & +26:24:11.45 & 0.316 & 0.040 & 0.279 & 0.047 & 0.450 & 0.072 & 10.2\\
  AMILA J025929+260944 & 02:59:29.75 & +26:09:44.99 & 0.230 & 0.038 & 0.265 & 0.052 &  &  & 10.3\\
  AMILA J025949+262518 & 02:59:49.91 & +26:25:18.08 & 0.470 & 0.058 & 0.432 & 0.083 & 0.561 & 0.111 & 10.9\\
  AMILA J030023+262604 & 03:00:23.02 & +26:26:04.58 & 0.638 & 0.070 & 0.605 & 0.074 & 0.756 & 0.118 & 11.3\\
  AMILA J025939+260555 & 02:59:39.93 & +26:05:55.91 & 0.392 & 0.064 & 0.382 & 0.081 & 0.422 & 0.086 & 11.3\\
  AMILA J030049+260644 & 03:00:49.39 & +26:06:44.21 & 0.651 & 0.049 & 0.659 & 0.050 & 0.614 & 0.137 & 12.5\\
  AMILA J025955+262726 & 02:59:55.12 & +26:27:26.03 & 8.925 & 0.090 & 9.191 & 0.090 & 8.467 & 0.149 & 12.5\\
  AMILA J025906+261529 & 02:59:06.92 & +26:15:29.86 & 0.431 & 0.044 & 0.437 & 0.055 &  &  & 13.8\\
  AMILA J025923+260551 & 02:59:23.40 & +26:05:51.17 & 0.422 & 0.062 & 0.411 & 0.062 &  &  & 13.8\\
  AMILA J025918+262340 & 02:59:18.18 & +26:23:40.19 & 0.279 & 0.058 & 0.276 & 0.057 &  &  & 14.1\\
  AMILA J025941+260219 & 02:59:41.06 & +26:02:19.90 & 1.806 & 0.066 & 1.775 & 0.064 & 2.065 & 0.160 & 14.3\\
  AMILA J030112+261823 & 03:01:12.77 & +26:18:23.39 & 0.279 & 0.056 & 0.257 & 0.055 &  &  & 14.7\\
  AMILA J025953+263010 & 02:59:53.73 & +26:30:10.76 & 0.300 & 0.055 & 0.304 & 0.059 &  &  & 15.3\\
  AMILA J030009+263101 & 03:00:09.92 & +26:31:01.28 & 1.045 & 0.058 & 1.034 & 0.068 & 1.296 & 0.309 & 15.8\\
  AMILA J030057+262650 & 03:00:57.89 & +26:26:50.90 & 0.447 & 0.040 & 0.460 & 0.044 &  &  & 16.0\\
  AMILA J030115+260843 & 03:01:15.49 & +26:08:43.98 & 0.272 & 0.060 & 0.271 & 0.061 &  &  & 16.4\\
  AMILA J030128+261639 & 03:01:28.35 & +26:16:39.90 & 0.434 & 0.087 & 0.426 & 0.090 &  &  & 17.9\\
  AMILA J025949+263246 & 02:59:49.66 & +26:32:46.80 & 0.269 & 0.066 & 0.269 & 0.066 &  &  & 18.0\\
  AMILA J030016+263346 & 03:00:16.53 & +26:33:46.54 & 1.253 & 0.069 & 1.253 & 0.069 &  &  & 18.6\\
  AMILA J025857+262447 & 02:58:57.33 & +26:24:47.79 & 0.924 & 0.063 & 0.924 & 0.063 &  &  & 18.6\\
  AMILA J030118+260504 & 03:01:18.68 & +26:05:04.94 & 0.621 & 0.074 & 0.621 & 0.074 &  &  & 18.7\\
  AMILA J030133+261322 & 03:01:33.86 & +26:13:22.22 & 0.759 & 0.075 & 0.759 & 0.079 &  &  & 19.2\\
  AMILA J030118+260352 & 03:01:18.83 & +26:03:52.90 & 0.672 & 0.067 & 0.672 & 0.067 &  &  & 19.4\\
  AMILA J030128+260657 & 03:01:28.24 & +26:06:57.29 & 0.352 & 0.066 & 0.352 & 0.066 &  &  & 19.7\\
  AMILA J030132+262143 & 03:01:32.00 & +26:21:43.76 & 0.541 & 0.081 & 0.541 & 0.081 &  &  & 19.8\\
  AMILA J030035+263425 & 03:00:35.48 & +26:34:25.11 & 2.565 & 0.065 & 2.565 & 0.065 &  &  & 20.1\\
  AMILA J030138+261919 & 03:01:38.62 & +26:19:19.75 & 0.445 & 0.073 & 0.445 & 0.073 &  &  & 20.6\\
  AMILA J025910+263125 & 02:59:10.53 & +26:31:25.00 & 0.541 & 0.068 & 0.541 & 0.068 &  &  & 20.7\\
  AMILA J030112+263056 & 03:01:12.55 & +26:30:56.11 & 0.690 & 0.066 & 0.690 & 0.066 &  &  & 21.2\\
  AMILA J030059+255646 & 03:00:59.08 & +25:56:46.12 & 1.245 & 0.073 & 1.245 & 0.073 &  &  & 21.7\\
\hline
\end{tabular}

\end{table*}

\begin{figure}
  \begin{center}
    \includegraphics[width=\linewidth]{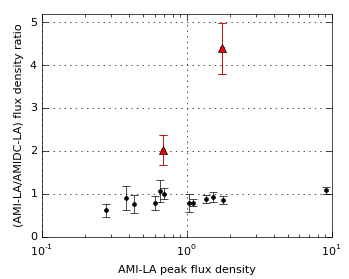}
    \caption{Flux density ratios for compact sources detected in both the \AMIAC-LA survey data and new \AMIDC-LA data.  Error bars include local thermal noise estimates as well as a 5\% calibration error.  Only two sources have varied significantly; these are plotted in red and with triangle markers.}
    \label{Fi:LA_fluxes}
  \end{center}
\end{figure}

For the fourteen sources that are detected in both maps, we compare the flux densities to check for significant variation.  The flux ratios are plotted in Fig.~\ref{Fi:LA_fluxes}, where the error bars include 5\% calibration uncertainties added in quadrature with the local thermal noise estimates.  Only two of these sources (AMILA J030001+262059 and AMILA J025935+261727) have varied significantly (i.e.\ $|S_{\mathrm{AMI_{AC}}}-S_{\mathrm{AMI_{DC}}}|/\bar{S}>3\sigma$, where $\bar{S}$ is the mean of the two flux densities).

The survey map and inner 19 pointings of the \AMIDC\ map have similar noise levels of $\approx\,30$ -- $40\,\upmu$Jy\,beam$^{-1}$ so we expect the same sources to be detected in this region; in fact five common sources are detected in both.  Three sources are detected in the \AMIDC\ map and not in the \AMIAC\ map.  One of these (AMILA J030032+261849) is next to a brighter source and is not detected due to a combination of the poorer dynamic range in the \AMIAC\ map and a slightly reduced flux density; we made a manual fit to the nearby sources using \textsc{jmfit} and obtained a flux density of $200 \pm 60 \upmu$Jy\,beam$^{-1}$, consistent with the \AMIDC\ flux within the noise levels.  The second and third (AMILA J025955+260842 and AMILA J025936+261343) are just visible at $\approx\,3.7 \sigma$ and 3.0$\sigma$ respectively in the \AMIAC\ map and also have consistent flux densities given the noise levels.   One source (AMILA J030010+261202) is detected in the \AMIAC\ map and not in the \AMIDC\ map; although it should have been detected at 6$\sigma$, there is no trace of it in the map and is probably a variable source caught at higher flux density during the previous observations.  The situation is similar in the outer region of the \AMIDC\ map, where all sources expected to be detected based on the higher noise level of $\approx$\,100\,$\upmu$Jy\,beam$^{-1}$ are detected; sources at just under the detection limit are visible in the \AMIDC\ map.

We therefore have confidence that the overall source environment has not changed significantly between the two sets of observations, and combine both sets of data to reduce the noise level and detect as many sources as possible, while suppressing artefacts in the \AMIAC\ data.  We average in the map plane since the survey and follow-up pointing centres do not coincide, using the noise maps generated by \textsc{source\_find} as weights for the average.  We do not attempt any correction for the small frequency shift since we will allow for small changes in the source flux density due to calibration offsets and/or variability when modelling the sources.  The combined LA map with the positions of the source detections is shown in Fig.~\ref{Fi:comb_map}.

\subsection{SA data comparison}

\begin{figure*}
  \centerline{\hbox to 10.5cm{\AMIAC-SA \hss \AMIDC-SA\hspace{1cm}}}
  \begin{center}
    \includegraphics[trim={0cm 1.0cm 3.4cm 1.5cm},clip=,height=0.33\textheight]{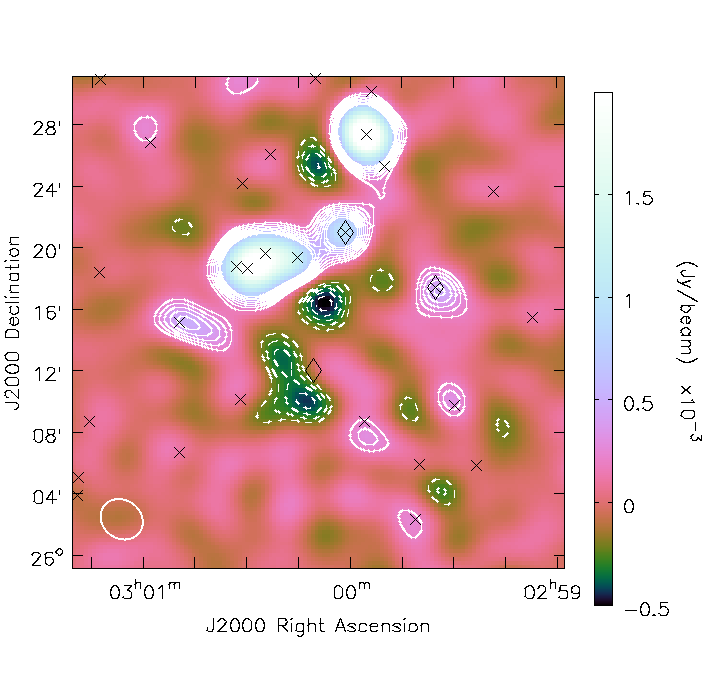}\includegraphics[trim={0cm 1.0cm 0cm 1.5cm},clip=,height=0.33\textheight]{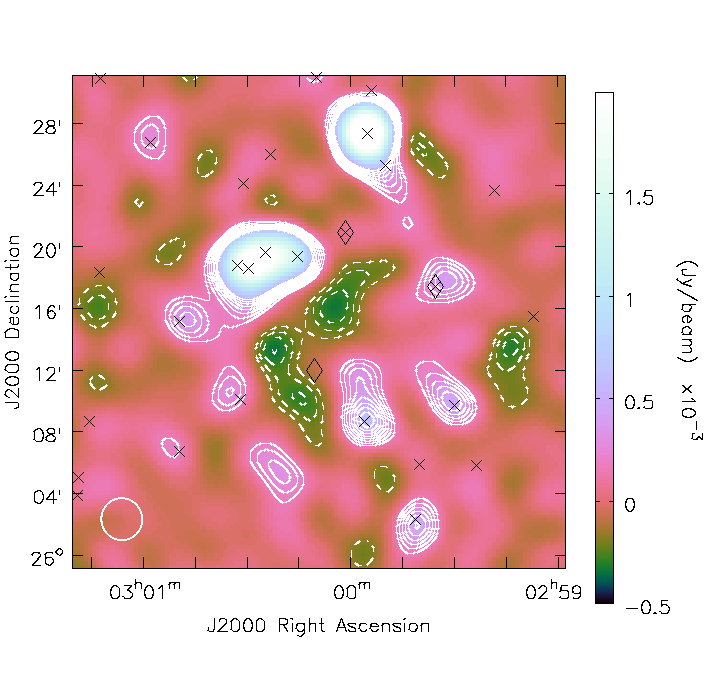}
    \caption{\AMIAC-SA (left) and \AMIDC-SA (right) maps of AMI-CL~J0300+2613.  The colour-scale is the same in both maps and is truncated to show low-surface-brightness features.  White contours (dashed for negative) are overlaid at $\pm 3, 4, 5, ... 10\sigma$, where $\sigma$ is the thermal noise measured on the respective maps as given in Table~\ref{tab:AMI_obs}.  The white ellipse in the left-hand corner of each map shows the synthesised beam, and the crosses and diamonds show the LA source positions as in Fig.~\ref{Fi:comb_map}.}
    \label{Fi:SA_maps}
  \end{center}
\end{figure*}

We first make a qualitative comparison of the \AMIAC-SA and \AMIDC-SA maps.  In both cases, \textsc{clean} was run blindly, with no boxes set to influence the choice of \textsc{clean} components, to a threshold of 3\,$\times$ the noise on the dirty map.  Natural weighting was used.  The two maps are shown in Fig.~\ref{Fi:SA_maps}.  The maps both show a decrement at the centre, with an extension to the south-east; however, the central decrement is deeper in the \AMIAC\ map at $\approx$\,500\,$\upmu$Jy\,beam$^{-1}$ ($\approx\,7.5 \sigma$) compared to $\approx$\,300\,$\upmu$Jy\,beam$^{-1}$ ($\approx\,6 \sigma$) in the \AMIDC\ map.  While the brightness of extended emission in an interferometric map depends on the inclusion and relative weighting of the short baselines present, both datasets have very similar $uv$-plane coverage (and the same physical baselines) so this should not cause the difference.  The other noticeable difference between the two maps is the reduced flux density of the compact source to the north of the decrement; this is the variable source AMILA J030001+262059 identified in Section~\ref{S:point_sources}.

\subsection{Cluster analysis}

We analysed both datasets using the cluster analysis software package \textsc{McAdam} \citep{2009MNRAS.398.2049F}; this fits simultaneously for cluster and compact source parameters while taking into account instrumental noise, primary CMB anisotropies, and confusion from radio sources below the LA detection threshold, in a Bayesian manner using the nested sampling algorithm \textsc{MultiNest} \citep{2009MNRAS.398.1601F}.  For the cluster model, we used a Navarro-Frenk-White (NFW; \citealt{1997ApJ...490..493N}) dark matter profile in hydrostatic equilibrium with a gas pressure distribution described by a generalised NFW (GNFW; \citealt{2007ApJ...668....1N}) profile with the shape parameters given in \citet{2010A&A...517A..92A}; for more details of the model see \citet{2012MNRAS.423.1534O}.  We refer to this as the DM-GNFW model.  We imposed a joint mass-redshift prior based on the cluster number counts of \citet{2008ApJ...688..709T} and fixed the gas mass fraction at $r_{200}$ to 0.13 \citep{2011ApJS..192...18K}.  We set a fairly tight prior on the position of the cluster (a Gaussian with $\sigma = 1$\,arcmin from the peak of the central decrement visible on the map), to concentrate the analysis on the central decrement.  

Each radio source has its position fixed to that determined by the LA.  Sources with flux density $>4\sigma$, where $\sigma$ is the noise value on the respective SA map, have their flux density $S$ and spectral index fitted, where the prior on the flux density is Gaussian with a 20\% width to account for inter-array calibration uncertainty and possible variability, and the prior on the spectral index is based on the 9C 15 -- 22\,GHz spectral index distribution \citep{2007MNRAS.379.1442W}.  Sources with flux density $<4\sigma$ have their flux densities fixed to the LA values, and spectral indices fixed to values determined from \citet{2013MNRAS.429.2080W} to be the median of the spectral index distribution at the appropriate flux density.  The flux-density priors for the three variable sources (if fitted) are centred at the appropriate value for the epoch and have a wider 40\% width since the SA and LA data were not necessarily taken at exactly the same time; all others are as determined from the combined LA map.  The parameters and priors on each are summarised in Table~\ref{tab:priors}.

\begin{table}
\centering
\caption{Summary of priors used in the Bayesian cluster analysis.  The top group of parameters relates to the cluster model while the bottom relates to the radio point sources. `NV' and `V' refer to non-variable and variable point sources, respectively.}\label{tab:priors}
\begin{tabular}{lcc}\hline
Parameter & Prior type & Limits \\ \hline
$x_{0}$, $y_{0}$ & $\mathcal{N}(\upmu = \mathrm{map\: peak},$ & \\
 & $\sigma = 1$\,arcmin) & \\
$z$ & Tinker($z$,$M_{200}$) & [0.2, 2] \\
$M_{200}$ & Tinker($z$,$M_{200}$) & [1, 60] $\times 10^{14} M_{\odot}$ \\
$f_{\mathrm{gas},200}$ & $\delta(0.13)$ & \\ \hline
$x_{s,i}$, $y_{s,i}$ & $\delta(\mathrm{LA})$ & \\
$S_{i}$ ($S_{i}>4\sigma_{\mathrm{SA}}$, NV) & $\mathcal{N}(\upmu=S_{i,\mathrm{LA}},$ & \\
 & $\sigma = 0.2 \times S_{i,\mathrm{LA}})$ & [0, $\inf$) \\
$S_{i}$ ($S_{i}>4\sigma_{\mathrm{SA}}$, V) & $\mathcal{N}(\upmu=S_{i,\mathrm{LA}},$ & \\
 & $\sigma = 0.4 \times S_{i,\mathrm{LA}})$ & [0, $\inf$) \\
$\alpha_{i}$ ($S_{i}>4\sigma_{\mathrm{SA}}$) & 9C & \\
$S_{i}$ ($S_{i}<4\sigma_{\mathrm{SA}}$) & $\delta(S_{i,\mathrm{LA}})$ &  \\
$\alpha_{i}$ ($S_{i}<4\sigma_{\mathrm{SA}}$) & $\delta(\alpha(S_{i,\mathrm{LA}}))$ & \\ \hline
\end{tabular}
\end{table}

\begin{figure}
  \centerline{\hbox to 0.8\linewidth{DM-GNFW model \hss Isothermal $\beta$ model}}
  \begin{center}
    \includegraphics[height=0.21\textheight]{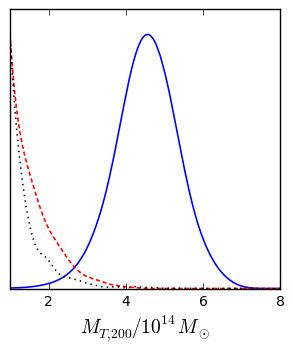}\includegraphics[height=0.21\textheight]{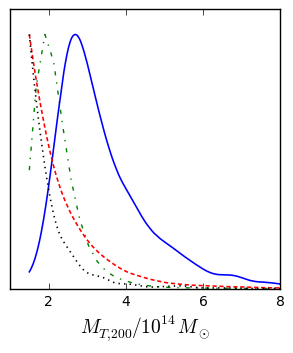}
    \caption{Left: marginalised 1-D mass posteriors as derived from \AMIAC\ data (blue solid line), \AMIDC\ data (red dashed line), and the prior distribution (black dotted line), using the DM-GNFW model.  Right: mass posteriors and prior as derived using the $\beta$-model, with the addition of the CARMA posterior (green dot-dashed line). Note that the $x$-axis scales are the same for ease of comparison, but the prior lower limit for the $\beta$-model analysis was set to $1.5 \times 10^{14} M_{\odot}$.}
    \label{Fi:mass_posteriors}
  \end{center}
\end{figure}

For each dataset, we ran our Bayesian analysis software with a model consisting of cluster and point sources (the `cluster' run), and with point sources only (the `null' run).  The ratio between the Bayesian evidences for these two runs can be used for model selection, i.e.\ to quantify whether the data are more consistent with or without a cluster being present.  The \AMIAC\ data had an evidence ratio of e$^{9.9}$, showing significant evidence for a cluster, while the \AMIDC\ data had an evidence ratio of e$^{1.4}$, showing only marginal evidence for the presence of a cluster.  Along with the difference in evidence ratios, it can also be seen that the mass posteriors are discrepant; the \AMIAC\ posterior puts a definite constraint on the mass at $M_{T,200} =  (4.54 \pm 0.83) \times 10^{14} M_{\odot}$ while the \AMIDC\ posterior can only provide an upper limit, $M_{T,200} < 1.79 \times 10^{14} M_{\odot}$; the marginalised mass posteriors are shown, together with the prior, in Fig.~\ref{Fi:mass_posteriors}.  Such a different result, for qualitatively similar maps, can be understood if the shape of the decrement in the \AMIDC\ data in $uv$-space does not agree well with the model.  In this case the evidence ratio is decreased, and the posterior on mass becomes dominated by the prior (which strongly prefers low mass) rather than the likelihood, so a much lower mass is preferred even though the decrement in the \AMIDC\ map is comparable to the decrement in the \AMIAC\ map.

We next test the robustness of this result by making various changes to the cluster priors and model, including: allowing the cluster to have an ellipsoidal geometry on the plane of the sky; setting a wider positional prior so that the extended `tail' of the decrement to the South is also found by the sampler; increasing the lower limit of the mass prior and using a \citet{2001MNRAS.321..372J} mass prior so that lower masses are not so strongly preferred by the prior; and changing the model to a purely phenomenological description of the SZ decrement (see \citealt{2012MNRAS.423.1463A} for details of this model and the parameter priors).  Each case is consistent in the general result that the evidence for the presence of a cluster in the \AMIDC\ data is reduced compared to the \AMIAC\ data.  The DM-GNFW models always indicate a reduced mass, and the phenomenological models always indicate a much more extended decrement (so that more cluster signal is resolved out) and a less negative temperature.  The phenomenological models give more significant evidence for the presence of a decrement in the \AMIDC\ data (but always much lower evidence values than when used with the \AMIAC\ data).  

\begin{figure*}
   \centerline{\hbox to 10.5cm{\AMIAC-SA \hss \AMIDC-SA\hspace{1cm}}}
    \centering
    \includegraphics[trim={0cm 1.0cm 3.4cm 1.5cm},clip=,height=0.33\textheight]{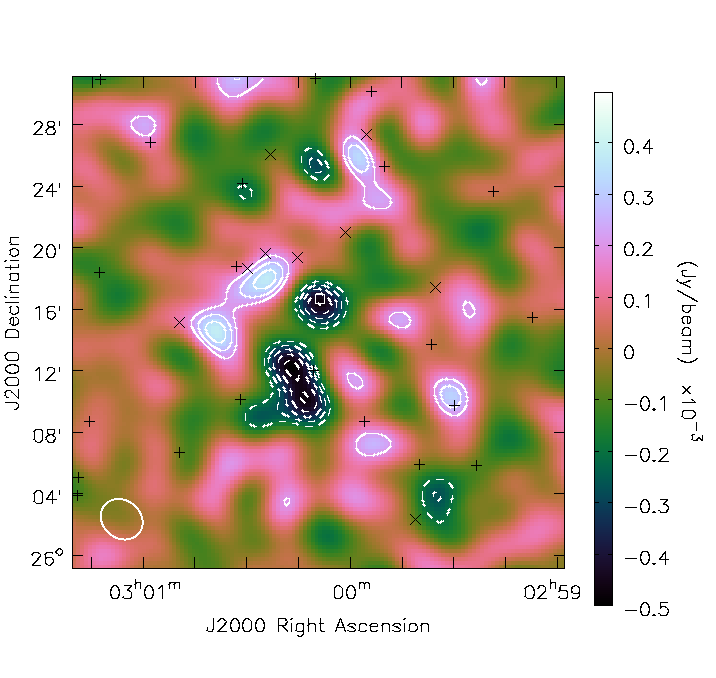}\includegraphics[trim={0cm 1.0cm 0cm 1.5cm},clip=,height=0.33\textheight]{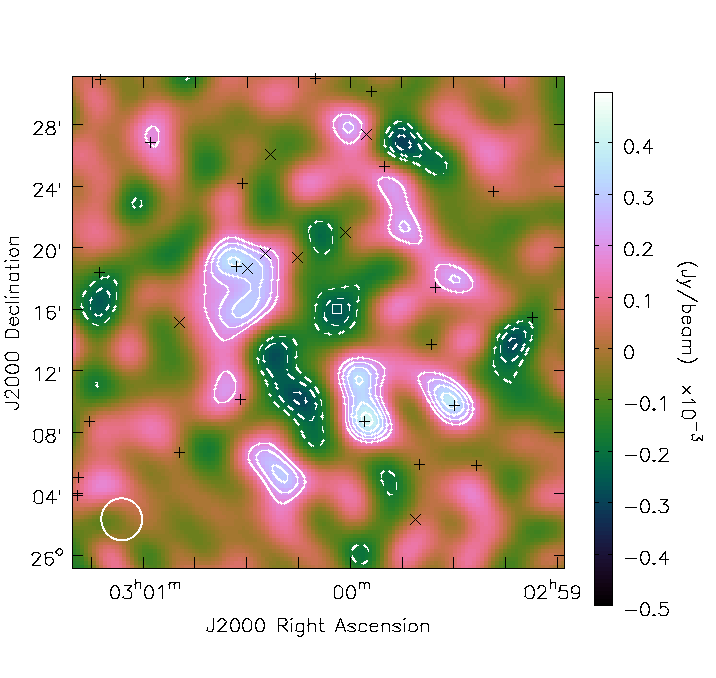}
    \caption{\AMIAC-SA (left) and \AMIDC-SA (right) compact-source-subtracted maps of AMI-CL~J0300+2613.  The colour-scale is the same in both maps and is not truncated.  Contours are as in Fig.~\ref{Fi:SA_maps}.  The `$\times$' markers show the positions of sources for which the flux density and alpha were modelled simultaneously with the cluster parameters, while the `$+$' markers show the positions of less significant sources which were subtracted using the LA source parameter estimates.  The small box shows the \textsc{McAdam} estimate of the cluster centre.  The white ellipse in the bottom left-hand corner shows the synthesised beam.}
    \label{Fi:SA_maps_sub}
\end{figure*}

For the \AMIDC\ data, since there are significant positive residuals present at the location of some sources on the source-subtracted map (see Fig.~\ref{Fi:SA_maps_sub}), we also investigated widening the flux density priors and allowing the positions of some sources to shift slightly, to remove as much positive emission as possible; again, the general result is unchanged.

Since the CARMA analysis was performed using an isothermal $\beta$ model for the cluster gas distribution with priors as listed in \citet{2012MNRAS.423.1463A}, we also ran the \AMIAC\ and \AMIDC\ analysis using the same model and priors for a fair comparison.  The resulting mass posteriors, along with those for the CARMA data, are also plotted in Fig.~\ref{Fi:mass_posteriors}, and it can be seen that the \AMIDC\ posteriors are in much better agreement with the CARMA posteriors than the \AMIAC\ posteriors.

\begin{figure}
  \centerline{GLEAM 200\,MHz\hspace{1cm}}
  \begin{center}
    \includegraphics[trim={0cm 0cm 0cm 0.7cm},clip=,width=\linewidth]{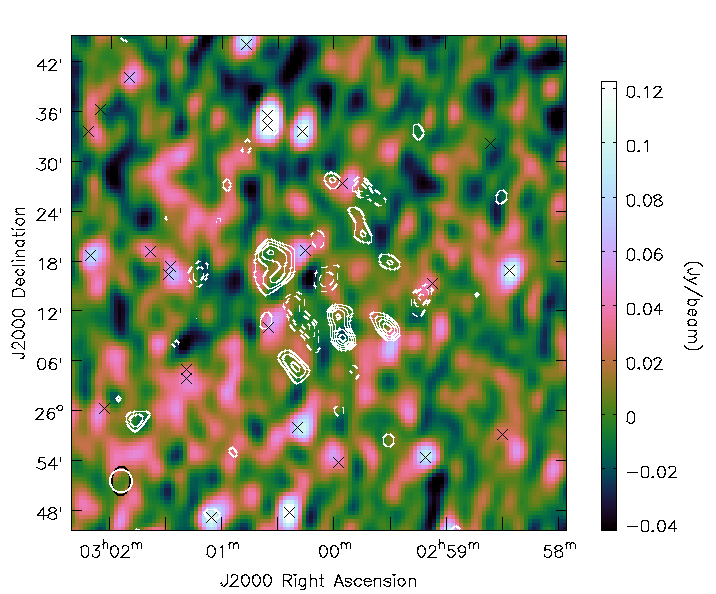}
    \caption{The GLEAM 200\,MHz map (colour-scale), overlaid with the \AMIDC-SA source-subtracted contours (in white).  The crosses show the positions of low-frequency compact sources from the TIFR GMRT Sky Survey Alternative Data Release (TGSS ADR1; \citealt{tgss} catalogue.  The colour-scale is truncated to show any low-surface-brightness extended features; contours are as in Fig.~\ref{Fi:SA_maps}.  The white and black ellipses superposed in the bottom left-hand corner show the \AMIDC-SA and GLEAM synthesised beams, respectively.}
    \label{Fi:GLEAM}
  \end{center}
\end{figure}

In Fig.~\ref{Fi:SA_maps_sub} we show the point-source-subtracted \AMIAC\ and \AMIDC\ maps, using the source parameters as fitted simultaneously with the cluster parameters (using the DM-GNFW model and the 1\,arcmin prior on cluster position).  There are significant positive residuals in both maps, indicating the presence of positive extended emission which was not detected on the LA map.  This could be a radio relic, which is a region of synchrotron emission caused by acceleration of relativistic electrons by shocks caused by cluster mergers; these are steep spectrum sources but have been detected at 15\,GHz (e.g.\ \citealt{2014MNRAS.441L..41S}; \citealt{2016MNRAS.455.2402S}), or indeed steep-spectrum synchrotron emission resulting from radio-jet activity.  We have checked the GaLactic and Extragalactic All-sky MWA (GLEAM) survey \citep{2017MNRAS.464.1146H}, at $\approx$\,200\,MHz and with good sensitivity to extended structures, and find no trace of the extended emission (see Fig.~\ref{Fi:GLEAM}).  \citet{2016MNRAS.455.2402S} fitted broken power-law spectra to the integrated flux densities of the two relics that have been detected at 15\,GHz, the `Sausage' and `Toothbrush' relics.  Using these fits, the relic should be either 450$\times$ or 240$\times$ brighter at 200\,MHz than at 15\,GHz, respectively.  The synthesised beam on the GLEAM map is the same size as the \AMIDC-SA beam to within 4\%, and we measure the noise level to be 18.5\,mJy\,beam$^{-1}$ using the \textsc{imean} task in \textsc{aips}.  The surface brightness of the positive emission ranges from $\approx$\,300 -- 650\,$\upmu$Jy\,beam$^{-1}$ on the \AMIDC-SA primary-beam-corrected map in the case that we widen the source priors to subtract as much positive emission as possible, so even in the `Toothbrush' case the faintest emission should be visible at $\approx$\,4$\sigma$ on the GLEAM map.  We therefore conclude that the positive emission is unlikely to be a radio relic or other form of synchrotron emission.

Given the reduced evidence for the presence of a cluster, we search for alternative explanations for the emission in datasets available at other wavebands.

\subsection{Dust emission}\label{S:dust}

\begin{figure}
  \centerline{\emph{Akari} WIDE-L (140\,$\upmu$m)\hspace{0.5cm}}
  \begin{center}
    \includegraphics[trim={1.5cm 1.5cm 0.6cm 0.5cm},clip=,width=\linewidth]{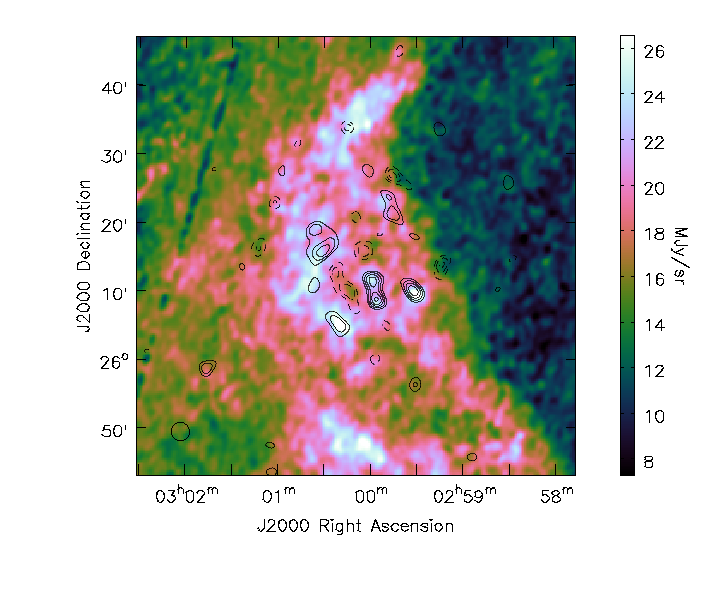}
    \caption{The \emph{Akari} WIDE-L ($140\,\upmu$m) map (colour-scale), overlaid with the \AMIDC-SA source-subtracted contours (in black).  The colour-scale is truncated to show low-surface-brightness extended features; contours are as in Fig.~\ref{Fi:SA_maps}.  The black ellipse in the bottom left-hand corner shows the \AMIDC-SA synthesised beam.}
    \label{Fi:AKARI}
  \end{center}
\end{figure}

Another source of extended positive emission at 15\,GHz is dust, either the tail of the greybody distribution or dust-correlated anomalous microwave emission (AME; first detected by \citealt{1997ApJ...486L..23L}).  We therefore compare the source-subtracted SA maps to the \emph{Planck} High Frequency Instrument (HFI) maps \citep{2016A&A...594A...8P}, the \emph{Akari} WIDE-L (140\, $\upmu$m) and WIDE-S (90\,$\upmu$m) band maps (\citealt{2007PASJ...59S.369M}; \citealt{2015PASJ...67...50D}), and the Wide-field Infrared Survey Explorer (\emph{WISE}, \citealt{2010AJ....140.1868W}) 12- and 25-$\upmu$m maps and find a clear correspondence between the AMI extended emission and a ring of emission visible in all the infrared/sub-mm maps mentioned, even though the AMI survey field was chosen to be well outside the Galactic plane at $\ell = 155.8^{\circ}, b = -28.3^{\circ}$.  Fig.~\ref{Fi:AKARI} shows the \AMIDC-SA source-subtracted contours overlaid on the \emph{Akari} WIDE-L-band map.  There is a clear correspondence between bright knots of emission in the filamentary dust structure and the AMI positive emission; the negative feature sits in the centre of the ring where there is less dust emission.

\begin{figure*}
  \centerline{\hbox to 15cm{\emph{Akari} WIDE-L (140\,$\upmu$m) \hss \emph{Akari} WIDE-S (90\,$\upmu$m)\hspace{0.5cm} \hss \emph{WISE} 12\,$\upmu$m\hspace{0.8cm}}}
  \begin{center}
    \includegraphics[trim={1.5cm 0cm 0.6cm 1cm},clip=,width=0.33\linewidth]{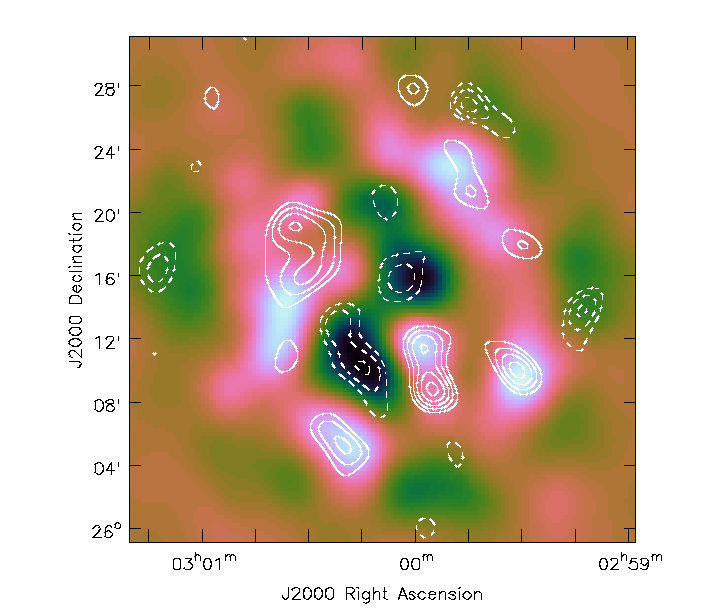}\includegraphics[trim={1.5cm 0cm 0.6cm 1cm},clip=,width=0.33\linewidth]{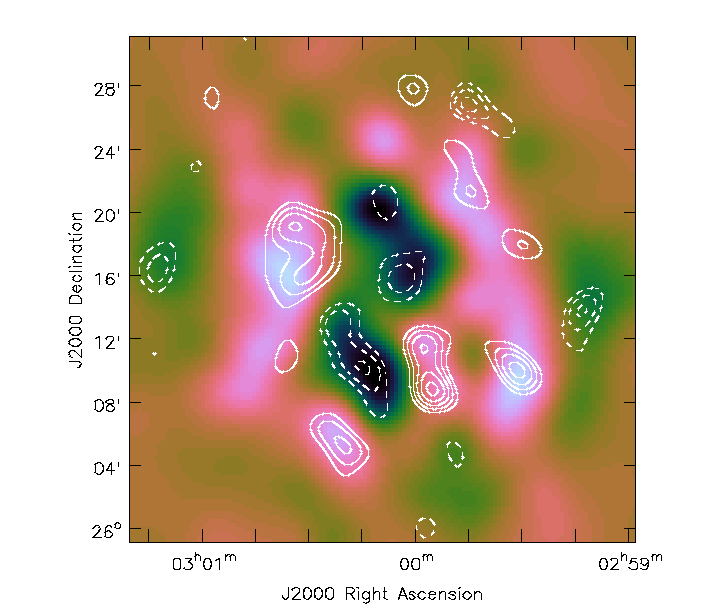}\includegraphics[trim={1.5cm 0cm 0.6cm 1cm},clip=,width=0.33\linewidth]{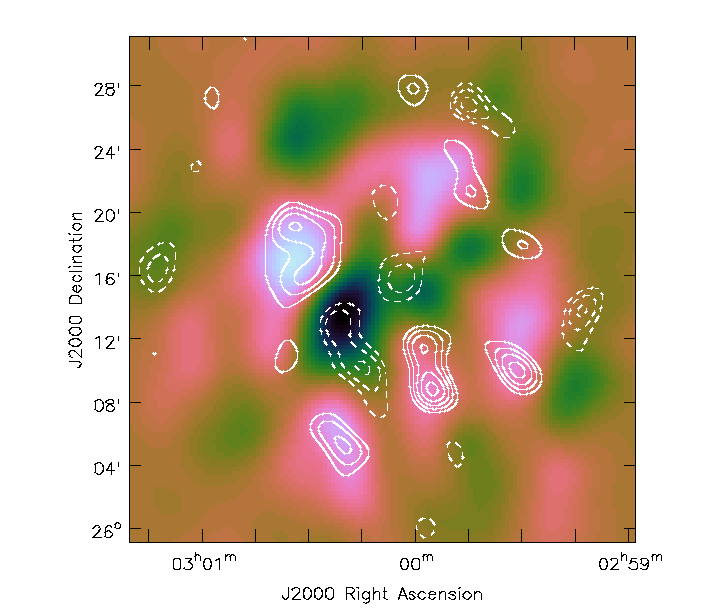}
    \caption{Simulated \AMIDC-SA observations of the \emph{Akari} WIDE-L (left) and WIDE-S (centre) band and \emph{WISE} 12 $\upmu$m (right) maps, with no added noise.  Colour scales are arbitrary; the zero level is orange.  The white contours show the \AMIDC-SA compact-source-subtracted residuals as in Fig.~\ref{Fi:SA_maps}.}
    \label{Fi:AKARI_sim}
  \end{center}
\end{figure*}

Comparisons with interferometric observations of resolved sources must take into account the spatial filtering applied by the interferometer.  This can be achieved by simulating interferometric observations of a sky model containing all relevant angular scales; i.e.\ at significantly higher angular resolution than the sky model and containing the large angular scales partially resolved out by the interferometer.  To further test the apparent AMI-dust correspondence, we simulated AMI observations of the \emph{Akari} WIDE-L- and WIDE-S-band and the \citet{2014ApJ...781....5M} source-subtracted \emph{WISE} 12\,$\upmu$m maps, which have angular resolutions 88, 78 and 15\,arcsec respectively.  The \emph{Planck} HFI maps are too low-resolution for this procedure and the \emph{WISE} 25-$\upmu$m maps are complicated by the presence of point sources.  We used the same $uv$-plane sampling as in the real \AMIDC-SA observations and did not add any noise.  Since the $uv$-coordinates correspond to baseline length measured in $\lambda$, meaning that slightly different angular scales are sampled at different frequencies, we simulated eight frequency channels covering the AMI band with the same sky brightness (i.e.\ no spectral index correction was applied to the infrared maps).  Maps of these simulations, imaged in the same way as the \AMIDC-SA data, are shown in Fig.~\ref{Fi:AKARI_sim}.  We tested the apparent correspondence in both the $uv$- and map-planes: for each dust simulation, we calculated Pearson correlation coefficients between the simulated and the real (point-source-subtracted) \AMIDC-SA visibilities, and also between the simulated and real map pixels.  The correlation coefficients are listed in Table~\ref{tab:correlations}.  We note that the $uv$-plane-based $r$-values are quite low due to the low signal-to-noise on each visibility measurement, but due to the large number of measurements they are still significant as shown by the very low $p$-values (which indicate the probability of an uncorrelated system producing datasets with $r$-values at least as extreme as the calculated $r$-value).  We also note the natural tendency for the $r$-values to decline with frequency due to more of the extended emission being resolved out; this is also seen, for example, if we calculate $r$-values between the two \emph{Akari} simulations.  The `Channel 1' correlation coefficients tend to be lower, probably due to higher noise in this frequency bin.

\begin{table}
\centering
\caption{Pearson $r$-values and $p$-values (as calculated by the scipy.stats.pearsonr module) for \AMIDC-SA and simulated \emph{Akari} and \emph{WISE} observations, both in the map- and $uv$-plane.  For the $uv$-plane based correlations we quote values for the eight simulated channels.}\label{tab:correlations}
\begin{tabular}{lcccc}\hline
Type & AMI & Aux & $r$ & $p$ \\
 & channel & data &  & \\ \hline
Map & all & WIDE-L & 0.2748 & 0.00 \\
Map & all & WIDE-S & 0.2669 & 0.00 \\
Map & all & \emph{WISE}-12 & 0.2170 & 0.00 \\
\noalign{\vskip 1mm}
$uv$ & 1 & WIDE-L & 0.0078 & $4.10 \times 10^{-04}$ \\
$uv$ & 2 & WIDE-L & 0.0138 & $2.86 \times 10^{-10}$ \\
$uv$ & 3 & WIDE-L & 0.0145 & $3.30 \times 10^{-11}$ \\
$uv$ & 4 & WIDE-L & 0.0112 & $3.37 \times 10^{-07}$ \\
$uv$ & 5 & WIDE-L & 0.0101 & $2.57 \times 10^{-06}$ \\
$uv$ & 6 & WIDE-L & 0.0088 & $4.29 \times 10^{-05}$ \\
$uv$ & 7 & WIDE-L & 0.0069 & $1.28 \times 10^{-03}$ \\
$uv$ & 8 & WIDE-L & 0.0087 & $4.91 \times 10^{-05}$ \\
\noalign{\vskip 1mm}
$uv$ & 1 & WIDE-S & 0.0080 & $3.19 \times 10^{-04}$ \\
$uv$ & 2 & WIDE-S & 0.0148 & $1.57 \times 10^{-11}$ \\
$uv$ & 3 & WIDE-S & 0.0139 & $1.95 \times 10^{-10}$ \\
$uv$ & 4 & WIDE-S & 0.0116 & $1.04 \times 10^{-07}$ \\
$uv$ & 5 & WIDE-S & 0.0093 & $1.44 \times 10^{-05}$ \\
$uv$ & 6 & WIDE-S & 0.0072 & $8.20 \times 10^{-04}$ \\
$uv$ & 7 & WIDE-S & 0.0051 & $1.78 \times 10^{-02}$ \\
$uv$ & 8 & WIDE-S & 0.0099 & $3.82 \times 10^{-06}$ \\
\noalign{\vskip 1mm}
$uv$ & 1 & \emph{WISE}-12 & 0.0080 & $3.00 \times 10^{-04}$ \\
$uv$ & 2 & \emph{WISE}-12 & 0.0116 & $1.09 \times 10^{-07}$ \\
$uv$ & 3 & \emph{WISE}-12 & 0.0154 & $1.90 \times 10^{-12}$ \\
$uv$ & 4 & \emph{WISE}-12 & 0.0086 & $7.91 \times 10^{-05}$ \\
$uv$ & 5 & \emph{WISE}-12 & 0.0078 & $2.82 \times 10^{-04}$ \\
$uv$ & 6 & \emph{WISE}-12 & 0.0058 & $6.71 \times 10^{-03}$ \\
$uv$ & 7 & \emph{WISE}-12 & 0.0055 & $1.05 \times 10^{-02}$ \\
$uv$ & 8 & \emph{WISE}-12 & 0.0071 & $1.04 \times 10^{-03}$ \\
\hline
\end{tabular}
\end{table}

\subsection{Ring/decrement degeneracy}

\begin{figure}
  \begin{center}
    \includegraphics[width=\linewidth]{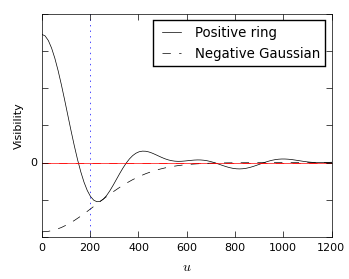}
    \caption{Real (black) and imaginary (red; insignificant due to symmetry) visibilities corresponding to a ring of positive emission and a negative Gaussian (see text for details), as a function of distance from the centre of the $uv$-plane, $u$, which corresponds to projected baseline length in units of $\lambda$.  The dotted blue vertical line shows the minimum projected SA baseline length.  The $y$-scale is arbitrary and the simulations have been normalised to have the same amplitude at $u = 250\lambda$.}
    \label{Fi:vis_comp}
  \end{center}
\end{figure}

To understand how a ring of positive emission can appear as a decrement, both visually in the map plane and in the $uv$-plane \textsc{McAdam} analysis (which only accounts for negative cluster emission and point sources), we performed some simple simulations.  We simulated a ring of emission of 6-arcmin thickness and a 6-arcmin inner radius, approximately mimicking the infrared emission, and a negative Gaussian with a 6-arcmin FWHM, approximating a cluster decrement.  In Fig.~\ref{Fi:vis_comp} we show the visibilities corresponding to these simulations (i.e.\ the simulation multiplied by the AMI-SA primary beam and Fourier transformed).  The `ring' visibilities have a negative real component on the same scale as the `cluster' decrement, and given the lack of baselines at $<200\lambda$ it would clearly be very difficult for the $uv$-plane analysis to distinguish between these two morphologies once noise is added, even with these very simple models.  Examining the more complicated Fourier transform of the \emph{Akari} maps, we see similar features.  This explains why the \textsc{McAdam} analysis marginally prefers the cluster model, even for the \AMIDC-SA data.

\begin{figure}
  \begin{center}
    \includegraphics[trim={0cm 0cm 0cm 0cm},clip=,width=0.49\linewidth]{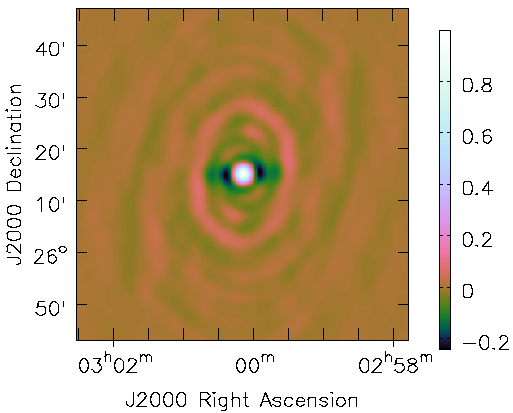}\includegraphics[width=0.51\linewidth]{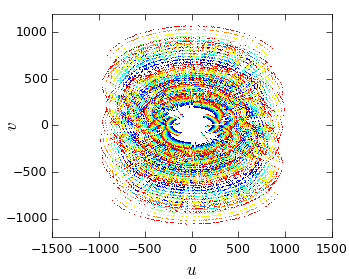}
    \caption{Dirty beam (left) and $uv$-coverage (right; i.e.\ projected baseline length in units of $\lambda$) for the \AMIDC-SA data.  Colours in the $uv$-coverage plot represent the 8 frequency bins; only every 120th point has been plotted for clarity.}
    \label{Fi:beam_uv}
  \end{center}
\end{figure}

\begin{figure}
  \begin{center}
    \includegraphics[trim={1cm 0cm 2cm 0cm},clip=,width=0.5\linewidth]{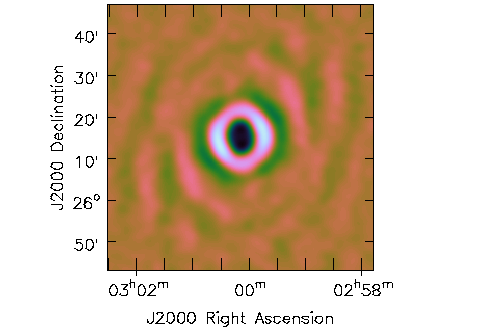}\includegraphics[trim={1cm 0cm 2cm 0cm},clip=,width=0.5\linewidth]{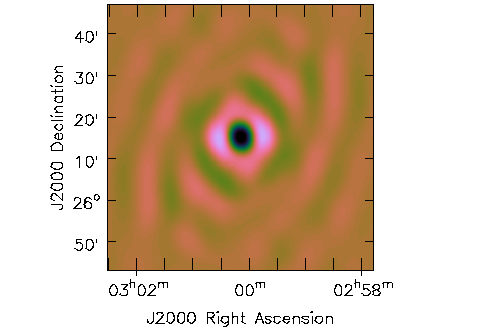}
    \caption{Simulated \AMIDC-SA dirty maps of a 4\,arcmin wide ring with a 4\,arcmin inner radius (left) and a negative Gaussian with a 6\,arcmin FWHM (right).  Colour scales are arbitrary; the zero level is orange.}
    \label{Fi:dirty_maps}
  \end{center}
\end{figure}

In the map plane, the degeneracy can be understood by considering the simulated dirty maps, which are the sky surface brightness convolved with the dirty beam, the Fourier transform of the $uv$-coverage.  Fig.~\ref{Fi:beam_uv} shows the dirty beam and $uv$-coverage for the \AMIDC-SA observations, and Fig.~\ref{Fi:dirty_maps} shows the dirty maps for the two simulations.  In this case the simulated ring has inner and outer radii of 4 and 8\,arcmin respectively, to better illustrate the problem.  The dirty beam has negative sidelobes of $\approx$\,25\% amplitude.  In the case of the ring, the negative sidelobes add up in the centre, producing a decrement of similar surface brightness to the positive ring.  The positive ring has brighter spots approximately aligned to the east--west axis; this is due to the ellipsoidal shape of the $uv$-coverage.  In the case of the decrement, the dirty map has positive bright spots in a similar place, which are produced by the negative beam sidelobes becoming positive when convolved with the decrement.  As in the $uv$-plane, it is very difficult to distinguish between these two morphologies without any additional information.

The higher significance of the decrement in the \AMIAC-SA map can also be explained as follows.  The analogue correlator had an imperfect point-source response which produced positive and negative ringing around sources.  If negative residuals from the positive emission in the ring happened to be at the right distance to add coherently in the same way as the dirty-beam sidelobes, this would produce an enhanced decrement, both in the map and the $uv$-plane analysis.  This also explains why the decrement remains significant in the \AMIAC-SA map after source subtraction -- idealised point-source subtraction does not remove the contribution from the artefacts.  From here on, we will concentrate on the \AMIDC-SA data. 

\subsection{Re-imaging the \AMIDC-SA data}\label{S:reimaging}

Since we now believe the decrement to be a misinterpretation of the interferometric measurement of the positive ring, we re-\textsc{clean}ed the \AMIDC-SA data interactively, placing \textsc{clean} boxes around the areas of positive emission rather than allowing \textsc{clean} components to be blindly placed in the most positive/negative regions.  This decreased the significance of both the central `decrement' and the more extended negative features to the south to $3\sigma$ or less, while increasing the significance of the positive features.

\subsection{Free--free analysis}

\begin{figure}
  \centerline{GB6 5\,GHz\hspace{1cm}}
  \begin{center}
    \includegraphics[trim={0cm 0.3cm 0.4cm 1.2cm},clip=,width=\linewidth]{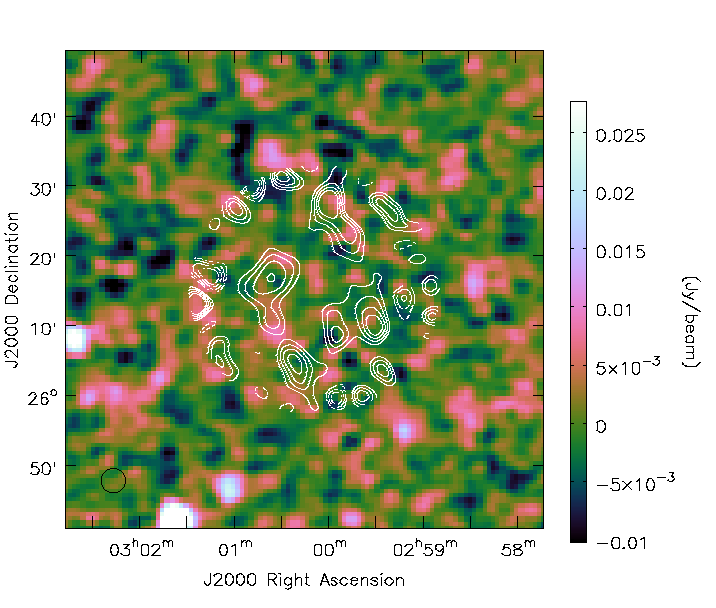}
    \caption{The GB6 (5\,GHz) map (colour-scale), overlaid with the \AMIDC-SA source-subtracted contours (in white), convolved to the GB6 resolution and extrapolated to 5\,GHz using $\alpha = 0.1$.  The colour-scale is truncated to show any low-surface-brightness extended features; contours are at ($\pm$4, 6, 8, 10) $\times$ 0.1\,mJy\,beam$^{-1}$.  Features at the edge of the \AMIDC\ primary beam are due to increased noise.  The black circle in the bottom left-hand corner shows the GB6 beam.}
    \label{Fi:GB6}
  \end{center}
\end{figure}

Extended, optically thin free--free emission could also account for the positive emission seen at 15\,GHz.  However, this is not a known star-forming region and checking the \citet{2014PASJ...66...17T} and \citet{2016MNRAS.458.3479M} young-stellar-object (YSO) catalogues we find only one YSO candidate (AllWISE~J030209.94+260045.9) nearby, well outside the AMI primary beam.  We therefore consider free--free unlikely to be the mechanism for the emission, but none-the-less check for visible emission at 5\,GHz in the GB6 survey map \citep{1994AJ....107.1829C}, which at resolution 3.5\,arcmin and containing angular scales up to $\approx$\,20\,arcmin has the correct spatial information.  We see no trace of the emission on the GB6 map; see Fig.~\ref{Fi:GB6}.  

We convolve the \AMIDC-SA map down to the GB6 resolution and extrapolate to 5\,GHz using the canonical optically-thin power-law index of $\alpha = 0.1$.   The maximum surface brightness on the extrapolated map is $\approx$\,1.0\,mJy\,beam$^{-1}$, while the GB6 map has a relatively high noise level of $\approx$\,3\,mJy\,beam$^{-1}$.  However, this is an upper limit given that the emission is clearly very resolved.  Using the simulated \AMIDC\ observation of the \emph{Akari} Wide-S map (see Section~\ref{S:dust}), which has the best correlation with the \AMIDC\ data, we estimate a scaling factor of $1.7 \times 10^{-4}$ to make the simulated \emph{Akari} visibilities consistent with the \AMIDC\ visibilities.  An estimate of the emission at 15\,GHz with all spatial scales present can therefore be made by scaling the \emph{Akari} map by this factor.  We then fit a twisted plane background to a polygon with edges surrounding the ring of emission, excluding the northern extension (see, e.g.\ \citealt{2007BASI...35...77G}).  We subtract this background to remove the largest scales which are not visible to GB6; convolve to the GB6 resolution and extrapolate to 5\,GHz using $\alpha = 0.1$.  In this case the surface brightness is $\approx\,7$ -- 10\,mJy\,beam$^{-1}$ and should be detectable by GB6.  While the correlation between the emission seen by AMI and \emph{Akari} is not perfect and so we cannot use this argument to conclusively rule out free--free as the origin of this emission, it seems unlikely.

\subsection{Greybody tail or AME?}

The \emph{Planck} 2015 data release \citep{2016A&A...594A...1P} included component-separated maps and fitted dust model parameter maps from several different methods.  We used the generalized needlet, internal linear combination (GNILC) \citep{2016A&A...596A.109P} dust parameter estimate maps to extrapolate the modified black-body emission fit to the AMI band and find that the expected thermal emission is at least $\approx\,20\,\times$ fainter than the observed AMI flux density, even before any spatial filtering of the extended emission is applied.  We therefore consider it very unlikely that the AMI emission is simply thermal emission and conclude that it is most likely to be AME.  We note that the \emph{Planck} component-separated AME map \citep{2016A&A...594A..10P} shows some structure in this region, which is $\approx\,10^{\circ}$ from the well-known AME region in Perseus (e.g.\ \citealt{2005ApJ...624L..89W}; \citealt{2013ApJ...768...98T}), but the sensitivity and angular resolution are both too low for a detection.  No emission is visible above the noise levels in the \emph{Planck} Low Frequency Instrument (LFI) maps, so we cannot construct a spectral energy distribution to check for the characteristic peak which would confirm the AME nature of the emission.

\section{Discussion}\label{S:discussion}

A strong contender for the origin of AME is electric dipole emission from rapidly rotating very small dust grains, with polycyclic aromatic hydrocarbons (PAHs) considered to be natural carriers of the emission due to their abundance and appropriate size (\citealt{1998ApJ...494L..19D}; \citealt{1998ApJ...508..157D}).  However, a definitive observational link between PAH abundance and AME has not been shown.  Some studies have shown greater correlations between 12-$\upmu$m emission, tracing the PAH abundance, than with longer wavelengths which trace the larger grains (e.g.\ \citealt{2006ApJ...639..951C}; \citealt{2010A&A...509L...1Y}) but the majority show no significant difference (e.g.\ \citealt{2011MNRAS.418.1889T}; \citealt{2014A&A...565A.103P}; \citealt{2016ApJ...827...45H}).  These AMI observations are consistent with the latter conclusion, since the 12-$\upmu$m simulation correlates slightly worse than the longer-wavelength simulations; however, none of the simulations is completely consistent with the AMI map, with parts of the emission (e.g.\ the eastern side of the ring in the \emph{Akari} maps, and the northern side of the ring in the \emph{WISE} map) visible in the infra-red yet not visible by AMI.

\begin{figure}
  \begin{center}
    \includegraphics[trim={1.5cm 1.5cm 0.6cm 1cm},clip=,width=\linewidth]{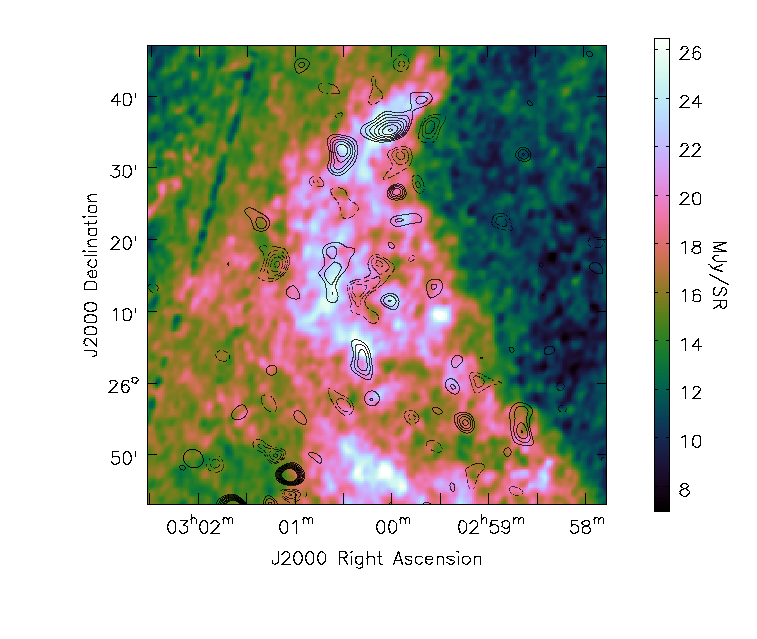}
    \caption{The \emph{Akari} Wide-L map of the region (colour-scale), overlaid with AMI-SA survey source-subtracted significance contours (in black).  Contours are at ($\pm$3, 4, 5, \dots, 10) $\times$ the local noise level on the map.  The black ellipse in the bottom left-hand corner shows the AMI-SA beam.}
    \label{Fi:survey}
  \end{center}
\end{figure}

This represents the only blind detection of AME on arcminute scales.  All previous blind detections (e.g.\ \citealt{1997ApJ...486L..23L}; \citealt{2005ApJ...624L..89W}; \citealt{2010A&A...509L...1Y}; \citealt{2016A&A...594A..10P}) have been at scales $>10$\,arcmin; higher-resolution detections have all been targeted observations of specific objects.  The AMI galaxy cluster survey can also be seen as a very deep survey for AME; we plan to reanalyse the rest of this survey field to search for additional positive extended structures and reobserve them with \AMIDC.  For example, it is clear that the bright northern extension of the ring is also seen in the AMI survey data (see Fig.~\ref{Fi:survey}).  More information is required to probe the nature of the AME in this field, including higher-frequency radio data to investigate the AME spectrum (we note that the CARMA observation did not contain enough short baselines to be useful in this regard), and high-resolution infra-red data in more bands to properly investigate the dust properties.  With more information, this survey could provide important clues as to the nature of AME given that the detection presented here does not appear to be associated with the usual AME-producers such as star-forming regions and dark clouds.

\section{Conclusions}\label{S:conclusions}

We have reobserved AMI-CL~J0300+2613, reported in \citet{2012MNRAS.423.1463A} to be a galaxy cluster detected via its SZ effect, with AMI equipped with a new digital correlator.  We find that:

\begin{enumerate}
\item{The SZ-decrement evidence for the presence of a cluster is much reduced in the \AMIDC\ data compared to the \AMIAC\ data, although the decrement is still visible in the map at lower significance.}
\item{By comparison with high-resolution sub-mm and infra-red maps that were not available at the time of the initial detection, we find that the apparent decrement is actually a misinterpretation of the interferometric measurement of a ring of dust-correlated emission.}
\item{Although we cannot entirely rule out free--free as the origin of the 15-GHz emission, we suggest that its origin is most likely to be Galactic AME, making it the first blind detection of AME on arcminute scales}
\item{Assuming the emission is AME, our analysis agrees with recent results that the AME does not necessarily correlate better with the 12-$\upmu$m emission which traces the PAH abundance.}
\item{We plan to reobserve other parts of the AMI blind cluster survey field to search for more AME from the structure visible in the infra-red maps.}
\end{enumerate}

\section*{Acknowledgments}

We thank the staff of the Mullard Radio Astronomy Observatory for their invaluable assistance in the commissioning and operation of AMI, which is supported by Cambridge and Oxford Universities.  We acknowledge support from the European Research Council under grant ERC-2012-StG-307215 LODESTONE.  We are grateful for IT knowledge exchange with the SKA project.  YCP acknowledges support from a Trinity College Junior Research Fellowship.  TMC, PJE, KJ and TZJ acknowledge STFC studentships.  TS acknowledges support from the ERC Advanced Investigator programme NewClusters 321271.  This research has made use of NASA's Astrophysics Data System Bibliographic Services. This publication makes use of data products from the Wide-field Infrared Survey Explorer, which is a joint project of the University of California, Los Angeles, and the Jet Propulsion Laboratory/California Institute of Technology, funded by the National Aeronautics and Space Administration.  This research is based on observations with \emph{AKARI}, a JAXA project with the participation of ESA.

\bsp \label{lastpage}

\end{document}